\documentclass[a4paper]{article}   

\usepackage{a4wide}
\usepackage{graphicx}          
\usepackage{hyphenat}
\usepackage{amsthm}
\theoremstyle{definition}
\newtheorem{defn}{Definition}
\newtheorem{thm}[defn]{Theorem}
\newtheorem{lem}[defn]{Lemma}
\newtheorem{exmp}[defn]{Example}

\usepackage{centernot,cancel}
\usepackage{amsmath} 
\usepackage{amssymb}  
\usepackage{xspace}

\usepackage{tikz}
\usetikzlibrary{automata,arrows,tikzmark,decorations.pathreplacing}
\usetikzlibrary {arrows.meta}
\usetikzlibrary{positioning}
\tikzset{initial text={},auto} 
\tikzstyle{every state}=[draw,shape=circle,inner sep=1pt,minimum size=12pt]

\usepackage{xspace}
\usepackage{hyperref}

\newcommand{\trans}[2]{\xrightarrow{#1}_{\! #2}}
\newcommand{\ntrans}[2]{\centernot{\xrightarrow{#1}}\!\!\!_{#2}}

\newcommand{\R}{\mathrel{R}}

\newcommand{\remove}[1]{}

\newcommand{\LC}{\mathrel{\mathbf{LC}}}
\newcommand{\AC}{\mathrel{\mathbf{AC}}}
\newcommand{\CR}{\mathrel{\mathbf{CR}}}

\newcommand{\FM}{\mathrel{\mathbf{FM}}}
\newcommand{\FMsubseteq}{\mathrel{\mathbf{FM}_{\subseteq}}}
\newcommand{\SC}{\mathrel{\mathbf{SC}}}
\newcommand{\SCn}{\mathrel{\mathbf{SC}^n}}
\newcommand{\PB}{\mathrel{\mathbf{PB}}}
\newcommand{\PBn}{\mathrel{\mathbf{PB}^n}}
\newcommand{\KT}{\mathrel{\mathbf{KT}}}
\newcommand{\incomparable}{\mathrel{\#}}

\newcommand{\aut}[1]{(\Q{#1},\allowbreak \Sigma,\allowbreak \trans{}{#1},\allowbreak \qinit{#1})}

\newcommand{\Q}[1]{Q_{#1}}
\newcommand{\qinit}[1]{q_{#1}^0}

\newcommand{\Sigmac}{\Sigma_c}
\newcommand{\Sigmau}{\Sigma_u}

\newcommand{\wrt}{w.r.t.\xspace}

\begin{document}

\title{Overview of  Controllability Definitions\\ in Supervisory Control Theory
} 
                              
\author{Jeroen J. A. Keiren \\
Formal System Analysis\\ 
Eindhoven University of Technology\\
\href{mailto:j.j.a.keiren@tue.nl}{j.j.a.keiren@tue.nl} \\
\and Michel A. Reniers\\
Control Systems Technology\\ 
Eindhoven University of Technology\\
\href{mailto:m.a.reniers@tue.nl}{m.a.reniers@tue.nl}
}

\maketitle
    
\begin{abstract}                          
In the field of supervisory control theory, the literature often proposes different definitions for the same concept, making it difficult to understand how these definitions are related.
This is definitely so for the fundamental notion of controllability of a supervisor \wrt\ a plant. 
This paper lists definitions of controllability found in the literature and studies their relationships in settings of both deterministic and nondeterministic automata.
In the general context, where both the supervisor and the plant are allowed to be nondeterministic, the notions of controllability as described by Flordal and Malik, and $\Sigmau$-admissibility by Kushi and Takai are equivalent. These are also the only notions that imply the traditional notion of (language) controllability.
From a practical perspective, one is often more interested in controllability of a supervised plant w.r.t. a plant. In this context, in addition to the previous two controllability notions, state controllability by Zhou et al.\ implies language controllability.
\end{abstract}

\section{Introduction}
\label{section:Introduction}

In the field of supervisory control theory, the basic problem addressed is the synthesis of a supervisor (or supervisory controller) from a model of an uncontrolled system, often referred to as a \emph{plant}, and a specification of the desired behavior of the supervised plant (that is, the plant in closed loop with the supervisor). An alternative formulation is often used where, instead of a specification of desired behavior, a set of requirements is provided that the supervised plant should satisfy.
In addition to the above-mentioned satisfaction of a specification or requirements, the synthesis procedure results in a supervisor that is maximally permissive, nonblocking and controllable~\cite{ramadge1989control}.

A fundamental notion in this problem statement is that of controllability. In supervisory control theory, the behavior of a plant is composed of controllable and uncontrollable events (coarsely representing actuation and sensing in the plant). Intuitively, controllability captures the assumption that a supervisor can only interfere with the controllable events of the plant.
Controllability is also of fundamental interest when the supervisor is not obtained by synthesis, but by means of another process, e.g., when it is manually crafted.

In the literature, different definitions for controllability have been proposed. It is sometimes hard to understand how these different notions relate and in which context they can be applied. To shed light on this, this paper lists the definitions of controllability found in the literature and studies their relationships.

Originally, controllability was defined as a relation between a specification language and a plant language~\cite{cassandras_introduction_2021,ramadge_control_1989}. To avoid ambiguity, this notion is called language controllability in the present paper.
The purpose is to guarantee that the plant language can be limited (by applying a supervisor) to achieve exactly the specification language by disabling only controllable events in the plant. 
In other works in the literature, controllability is also defined directly between a supervisor and a plant. In this case, the purpose is to establish that the supervisor does not disable uncontrollable events in the plant.
Finally, definitions of controllability that relate a supervised system and a plant are sometimes found. With language controllability in mind and assuming that the supervisor and the plant have the same set of events, controllability of a supervisor \wrt\ a plant is equivalent to controllability of the supervised plant \wrt\ that plant (see Lemma~\ref{lem:cc} in Section~\ref{subsec:lc}).

In the present paper, we prefer to represent both the supervisor and the plant using a (finite) automaton, and we prefer to formulate controllability as a property of a supervisor \wrt\ a plant. 

\paragraph*{Contribution(s)}
The contributions of this paper can be summarized as follows:
\begin{itemize}
\item We present an overview of the definitions of controllability from the literature and identify the contexts in which they are defined in terms of properties and assumptions on plant and supervisor (in Section~\ref{sec:controllability}). 
\item We identify the relations between controllability notions in different contexts of deterministic and nondeterministic plant and supervisor automata (in Section~\ref{sec:comparison}).
\item We reflect on the definitions in Section~\ref{sec:discussion}. We determine that only two other definitions imply language controllability in all identified contexts.
\item Considering the point of view that the controllability of a supervised plant \wrt\ a plant is the more relevant notion, we establish that three notions from literature are adequate under this interpretation.
\end{itemize}

The proofs of all theorems presented in this paper can be found in Appendix~\ref{sec:proofs}. In Section~\ref{sec:comparison}, examples are used as a means to prove that relations between notions of controllability do not exist. In addition, for these examples, a more detailed justification is given in Appendix~\ref{sec:proofs}.

\section{Related Work}
\label{sec:relatedwork}

In this paper, we discuss the fundamental notion of controllability in detail. In Table~\ref{tab:comparison} we give an overview of the different controllability notions that we analyse in this paper. We forego a detailed discussion of these notions of controllability here, as they will be studied in detail in Section~\ref{sec:controllability}.

\begin{table*}[htbp]
\newcommand{\lang}{L}
\newcommand{\deterministic}{D}
\newcommand{\nondeterministic}{N}
\caption{Overview of controllability notions. Columns $S$ and $P$ indicate whether supervisor and plant are  
deterministic ($\deterministic$) or nondeterministic ($\nondeterministic$) automata. Column $\subseteq$ indicates what assumption the corresponding definition makes on the inclusion of languages or automata.}
\label{tab:comparison}

\medskip
\begin{center}
\hrulefill

\scriptsize
\begin{tabular}{lll|ccc}
\textbf{Notion} & \textbf{Abbreviation} & \textbf{Definition} & $\mathbf{S}$ & $\mathbf{P}$ & $\mathbf{\subseteq}$ \\
\hline
Language controllability \cite{cassandras_introduction_2021,ramadge_control_1989} & $\LC$ & \ref{def:controllability} & $\nondeterministic$ & $\nondeterministic$ & \\
Control relation \cite{rutten_coalgebra_2000} & $\CR$ & \ref{def:control-relation} & $\deterministic$ & $\deterministic$ & \\
State controllability \cite{SvSR2008} & $\SC$ & \ref{def:state-controllability-su} & $\deterministic$ & $\nondeterministic$ & \\
Partial bisimulation \cite{rutten_coalgebra_2000} & $\PB$ & \ref{def:partial-bisimulation-det} & $\deterministic$ & $\deterministic$ & \\
Nondeterministic partial bisimulation \cite{baeten_partial_2010,markovski_employing_2015} & $\PBn$ & \ref{def:partial-bisimulation-nondet} & $\nondeterministic$ & $\nondeterministic$ & \\
Controllability for nondeterministic automata \cite{flordal_supervision_2006,flordal_compositional_2007} & $\FM$ & \ref{def:controllability-automata-flordal} & $\nondeterministic$ & $\nondeterministic$ & \\
$\Sigmau$-admissibility \cite{kushi_synthesis_2018} & $\KT$ & \ref{def:sigma-u-admissibility} & $\nondeterministic$ & $\nondeterministic$ & \\
Nondeterministic state controllability \cite{zhou_control_2006} & $\SCn$ & \ref{def:state-controllability-zhou} & $\nondeterministic$ & $\nondeterministic$ & $L(S) \subseteq L(P)$ \\
Controllability for nondeterministic subautomata \cite{flordal_supervision_2006,flordal_compositional_2007} & $\FMsubseteq$ & \ref{def:controllability-automata-flordal-subsupervisor} & $\nondeterministic$ & $\nondeterministic$ & $S \subseteq P$
\end{tabular}

\hrulefill
\end{center}
\end{table*}

For a context with both deterministic plant and supervisor, in~\cite{arnold_games_2003}, where a game-based approach for synthesis of (supervisory) controllers is discussed, a controller is always required to enable any uncontrollable event. In \cite{basu_quotient-based_2006} this approach is extended to nondeterministic plants and a state-dependent notion of (un)controllable events. In both works, controllability is captured as a modal $\mu$-calculus formula for which a satisfying model needs to be found.
In~\cite{reniers_validation_2024} a method is presented to express controllability in terms of modal $\mu$-calculus formulas such that controllability can be verified using model checking.

Controllability (as a property of a supervised system in comparison with an unsupervised plant) is also relevant in semantically richer contexts.
Typically, definitions in such contexts are generalizations of the definitions used in this paper.
In the setting with unobservable events, the definition of language controllability is essentially unchanged; see, e.g.,~\cite{cassandras_introduction_2021}.
Controllability in a setting with forcible events is considered in~\cite{golaszewski_control_1987,weidemann_optimal_2012}.
For timed discrete-event systems, controllability is extended in~\cite{brandin_supervisory_1994}. Generalizations to time-interval discrete-event systems and timed automata with forcible events are given in~\cite{brandin_supervisory_2024} and~\cite{rashidinejad_supervisory_2023}, respectively.

In some implementations of synthesis (such as in Supremica \cite{akesson2006supremica} and CIF \cite{fokkink_eclipse_2023}), specifications of admissible behavior by means of automata are dealt with by plantification  \cite{flordal_compositional_2007}. In plantification a requirement automaton is made complete by adding all missing uncontrollable transitions with a fresh state as the target state. Safe nonblocking supervisory control can in this way be reduced to nonblocking supervisory control. The result of plantification is controllable \wrt\ each of the notions of controllability studied in this paper.

\section{Languages and Automata}
\label{sec:languages}

Discrete-event systems are used to model systems. In this setting, we distinguish the \emph{plant}, i.e., the system to be controlled, and the \emph{supervisor}.
Plants are traditionally modeled using languages \cite{cassandras_introduction_2021,ramadge_control_1989}.
Let $\Sigma$ be a set of \emph{events}. Then $\Sigma^*$ denotes the set of all sequences of events. A language $L$ is a subset of $\Sigma^*$. We write $\epsilon$ for the empty sequence. 
Given $w = w_1w_2 \in \Sigma^*$, we call $w_1$ a \emph{prefix} of $w$.
For a language $L$, its \emph{prefix closure} $\overline{L}$ is defined as $\overline{L} = \{ w_1 \in \Sigma^* \mid \exists_{w_2 \in \Sigma^*}~w_1w_2 \in L \}$. If $L = \overline{L}$, we call $L$ \emph{prefix closed}. The concatenation $LL'$ of two languages is defined as $LL' = \{ w_1w_2 \in \Sigma^* \mid w_1 \in L \land w_2 \in L' \}$.

We partition the set of events $\Sigma$ into \emph{controllable} events $\Sigmac$ that can be influenced by the supervisor and \emph{uncontrollable} events $\Sigmau$ that the supervisor cannot affect.
Traditionally (e.g., in \cite{ramadge_control_1989}), a supervisor for plant $P \subseteq \Sigma^*$ is a function $f \colon P \to \Gamma$ such that $f(w) \in \Gamma$ prescribes the set of allowed events after $w$, which includes all uncontrollable events. 
Note that it is assumed that the event sets of the supervisor and the plant are the same. In this paper, the same assumption is made.
Formally, let $\Gamma \subseteq 2^\Sigma$ be a set of control inputs, with 
$\Sigmau \subseteq \gamma$ for all $\gamma \in \Gamma$. The supervised plant $P_f$ is defined inductively such that $\epsilon \in P_f$, and $wa \in P_f$ if $w \in P_f$, $a \in f(w)$ and $wa \in P$.

In practice, in particular when using tools, a finite description of languages is desired. Therefore, languages for the plant and the supervisor are often described using (finite) automata. This restricts the plant and the supervisor to regular languages. In some tools, such as Supremica \cite{akesson2006supremica} and CIF \cite{van2014cif,fokkink_eclipse_2023}, extended finite automata (i.e., automata with (finite domain) state variables) are used for representing plant and supervisor~\cite{skoldstam_modeling_2007}. In the remainder of this paper, we focus on plants and supervisors that are represented using automata.

In the context of supervisory control, an automaton is sometimes equipped with a finite set of marked states.
These primarily play a role when considering nonblockingness.
Since we are interested in studying controllability, and marked states do not play a role for that, we omit marked states from all definitions, even though some definitions of controllability in the literature take marked states into account. 

\begin{defn}
An \emph{automaton} $G$ is a tuple $\aut{G}$ where  
    $\Q{G}$ is a finite set of states; 
    $\Sigma_{G}$ is a finite set of events, which is partitioned into controllable events $\Sigmac$ and uncontrollable events $\Sigmau$;
    $\trans{}{G} {}\subseteq \Q{G} \times \Sigma_{G} \times \Q{G}$ is the transition relation;
    and
    $\qinit{G} \in \Q{G}$ is the initial state.
\end{defn}
\noindent
Note that the elements of the tuple are subscripted with the name of the automaton to easily identify which automaton is referred to. In case there can be no confusion these subscripts may be omitted.
In the remainder of this paper, when clear from the context, we leave the sets of controllable ($\Sigmac$) and uncontrollable actions ($\Sigmau$) implicit.
 
We typically use $a$ to denote arbitrary events, $c$ to denote arbitrary \emph{controllable} events, $u$, $u_1$, $u_2$, etc. to denote arbitrary \emph{uncontrollable} events, and $w$, $w_1$, $w_2$, etc. for sequences of events.
For $q,q' \in \Q{}$ and $a \in \Sigma$, we typically write $q \trans{a}{} q'$ instead of $(q,a,q') \in {} \trans{}{}$. We generalize this such that $q \trans{w}{} q'$ denotes the fact that there is a sequence of states connected by the events from the word $w$ starting from state $q$ and ending in state $q'$. We write $q \trans{a}{}$ (resp. $q \trans{w}{}$) to denote that there exists a $q' \in Q$ s.t.\ $q \trans{a}{} q'$ (resp. $q \trans{w}{} q'$).
Given a state $q \in Q$ and $a \in \Sigma$, we write $\Delta(q,a) = \{ q' \in Q \mid q \trans{a}{} q' \}$.

If for all $q,q',q'' \in \Q{}$ and $a \in \Sigma$, $q \trans{a}{} q'$ and $q \trans{a}{} q''$ implies $q' = q''$, we call the automaton \emph{deterministic}. If $G$ is deterministic and $q \trans{a}{}$, then $\Delta(q, a)$ is a singleton, and with slight abuse of notation, we write $\Delta(q, a)$ also for the single state that it contains.

Every automaton $G = \aut{}$ generates a discrete-event system in the traditional sense, i.e., as a prefix-closed language $L(G) = \{ w \in \Sigma^* \mid \qinit{} \trans{w}{} \}$ over alphabet $\Sigma$.

In this paper, automata are depicted graphically whenever appropriate. States are denoted by circles containing the name of the state. The initial state is indicated by a dangling incoming arrow. 
Solid (resp. dashed) arrows indicate transitions labelled by controllable (resp. uncontrollable) events.

The supervised plant is obtained as the synchronous product of plant and supervisor as defined below. 

\begin{defn}
\label{def:product}
The \emph{synchronous product} of automata $G = \aut{G}$ and $H = \aut{H}$, is the automaton $G \parallel H = \aut{G \parallel H}$, where
\begin{itemize}
    \item $\Q{G \parallel H} = \Q{G} \times \Q{H}$;
    \item 
    $\trans{}{G \parallel H}\, = \{ ((q_1,q_2),a, (q'_1,q'_2)) \mid q_1\trans{a}{G} q'_1, q_2 \trans{a}{H} q'_2 \}$;
    \item $\qinit{G \parallel H} = (\qinit{G},\qinit{H})$.
\end{itemize}
\end{defn}

Sometimes we may assume that a supervisor $S$ for a plant $P$ is a subautomaton of that plant as defined below.

\begin{defn}[Subautomaton]
Automaton $G = \aut{G}$ is a \emph{subautomaton} of automaton $H = \aut{H}$, denoted $G \subseteq H$, iff $\Q{G} \subseteq \Q{H}$, $\trans{}{G} {}  \subseteq {} \trans{}{H} {}\cap {} (\Q{G} \times \Sigma \times  \Q{H})$, and $\qinit{G} = \qinit{H}$. 
\end{defn}

\section{Definitions of Controllability in the Literature}
\label{sec:controllability}

In this section, we define the controllability notions mentioned in Table~\ref{tab:comparison}.
We present controllability notions as a binary relation between automata. For binary relations $R \subseteq \Q{S} \times \Q{P}$, we typically use infix notation $q \R q'$ instead of $(q, q') \in R$.

\subsection{Language Controllability}
\label{subsec:lc}

In the literature, a notion of controllability is given (on a specification language and a plant language) and a controllability theorem is provided that states that controllability of that specification is a sufficient condition for existence of a supervisor for realizing that specification~\cite{cassandras_introduction_2021,ramadge_control_1989}. To present all controllability notions in a single setting, we here consider the language of the supervisor as the specification.

When restricted to the languages of automata, this results in the following definition of controllability.
This formulation is based on the observation that languages generated by automata are prefix closed.
We refer to this notion as language controllability from now on to distinguish it from other notions of controllability defined later.

\begin{defn}[Language controllability]\label{def:controllability}
Supervisor language $L(S) \subseteq \Sigma^*$ is language controllable \wrt\ plant language $L(P) \subseteq \Sigma^*$ if $L(S) \Sigmau \cap L(P) \subseteq L(S)$.
We write $L(S) \LC L(P)$ iff $L(S)$ is language controllable \wrt $L(P)$.
\end{defn}
\noindent
Intuitively, if after a string $w$ in $L(S)$, an uncontrollable event $u$ is possible in $L(P)$, then $wu$ is also in the prefix closure of $L(S)$.

\begin{lem}
\label{lem:cc}
Let $S$ and $P$ be automata over the same event set. Then $L(S) \LC L(P)$ iff $L(S \parallel P) \LC L(P)$.
\end{lem}

\subsection{Control Relation}

In \cite{rutten_coalgebra_2000}, Rutten defines a \emph{control relation}~($\CR$) between two deterministic automata. This definition uses the fact that if an event is enabled in a certain state, then the target state is uniquely defined (due to the fact that the automaton is deterministic). 

\begin{defn}[Control relation]\label{def:control-relation}
Let $S = \aut{S}$ and $P=\aut{P}$ be deterministic automata with uncontrollable event set $\Sigmau$.
A relation $R \subseteq \Q{S} \times \Q{P}$ is a \emph{control relation} if, for all $q_S \in \Q{S}$ and $q_P \in \Q{P}$, $q_S \R q_P$ implies
\begin{itemize}
    \item for all $a \in \Sigma$ if $q_S \trans{a}{S}$ and $q_P \trans{a}{P}$, then $\Delta(q_S, a) \R \Delta(q_P, a)$;
    \item for all $u \in \Sigmau$ if $q_P \trans{u}{P}$, then $q_S \trans{u}{S}$ and $\Delta(q_S, u) \R \Delta(q_P, u)$.
\end{itemize}
We write $q \CR q'$ iff there exists a control relation $R$ such that $q \R q'$; and write $S \CR P$ if $\qinit{S} \CR \qinit{P}$.
\end{defn}

\subsection{State Controllability}

A notion of \emph{state controllability} is defined in \cite{SvSR2008}.
We here consolidate this notion in a single definition.

\begin{defn}[State controllability]\label{def:state-controllability-su}
Let $S = \aut{S}$ be a deterministic automaton and $P = \aut{P}$ be a (potentially nondeterministic) automaton.
Then $S$ is state controllable \wrt\ $P$, denoted $S \SC P$, iff
\begin{itemize}
    \item there exists an embedding $f \colon \Q{S} \to \Q{P}$ such that
    \begin{itemize}
        \item $f(\qinit{S}) = \qinit{P}$;
        \item for all $q, q' \in \Q{S}$ and $a \in \Sigma$, if $q \trans{a}{S} q'$ then $f(q) \trans{a}{P} f(q')$;
    \end{itemize}
    \item for all $q \in \Q{S}$, $F_S(q) \cap E_P(f(q)) \subseteq \Sigmac$.
\end{itemize}
where $E_i(q) = \{ a \in \Sigma \mid q' \in \Q{i} \land q \trans{a}{i} q' \}$ and $F_i(q) = \Sigma \setminus E_i(q)$.\footnote{The definition in~\cite{SvSR2008} furthermore considers marked states, and requires that marked states in $S$ are mapped to marked states in $P$.}
\end{defn}

\subsection{Partial Bisimilarity}

Rutten introduced partial bisimilarity in a coalgebraic setting to capture controllability of deterministic systems~\cite{rutten_coalgebra_2000}. 

\begin{defn}[Partial bisimulation]\label{def:partial-bisimulation-det}
Let
$S = \aut{S}$ and $P = \aut{P}$ be deterministic automata.
Relation $R \subseteq \Q{S} \times \Q{P}$ is a \emph{partial bisimulation} if, for for all $q_S \in \Q{S}$ and $q_P \in \Q{P}$, $q_S \R q_P$ implies
\begin{itemize}
    \item for all $a \in \Sigma$ if $q_S \trans{a}{S}$ then $q_P \trans{a}{P}$ and $\Delta(q_S, a) \R \Delta(q_P, a)$;
    \item for all $u \in \Sigmau$ if $q_P \trans{u}{P}$ then $q_S \trans{u}{S}$ and $\Delta(q_S, u) \R \Delta(q_P, u)$.
\end{itemize}
We write $q_S \PB q_P$ if there exists a partial bisimulation relation $R$ such that $q_S \R q_P$. We also call relation $\PB$ \emph{partial bisimilarity}.\footnote{\label{footnote-partial-bisim}The definitions in~\cite{baeten_partial_2010,markovski_employing_2015,rutten_coalgebra_2000} take marked states into account and additionally require that $q_S \R q_P$ implies $q_S$ is marked iff $q_P$ is marked.}
\end{defn}

Intuitively, the main difference between control relations and partial bisimulation is that a control relation allows the supervisor to execute controllable events that are not enabled in a related plant state. In partial bisimilarity this is not allowed. 

\subsection{Nondeterministic Partial Bisimilarity}
Partial bisimilarity was generalized to nondeterministic automata~\cite{baeten_partial_2010,markovski_employing_2015}.  In \cite{van_hulst_maximally_2017} partial bisimilarity is used to capture controllability in the context of synthesis of maximally permissive supervised plants for nondeterministic plants and (a subset of) modal logic specifications. The generalization to nondeterministic automata is as follows.

\begin{defn}{\normalfont \textbf{(Nondeterministic partial bisimu\-la\-tion)}}\label{def:partial-bisimulation-nondet}
Let
$S = \aut{S}$ and $P = \aut{P}$ be nondeterministic automata.
Relation $R \subseteq \Q{S} \times \Q{P}$ is a \emph{nondeterministic partial bisimulation} if, for all $q_S \in \Q{S}$ and $q_P \in \Q{P}$, $q_S \R q_P$ implies
\begin{itemize}
    \item for all $a \in \Sigma, q'_S \in \Q{S}$ such that $q_S \trans{a}{S} q'_S$, there exists $q'_P \in \Q{P}$ such that $q_P \trans{a}{P} q'_P$ and $q'_S \R q'_P$;
    \item for all $u \in \Sigmau, q'_P \in \Q{P}$ such that $q_P \trans{u}{P} q'_P$, there exists $q'_S$ such that $q_S \trans{u}{S} q'_S$ and $q'_S \R q'_P$.
\end{itemize}
We write $q_S \PBn q_P$ if there exists a nondeterministic partial bisimulation relation $R$ such that $q_S \R q_P$. This relation is called \emph{nondeterministic partial bisimilarity}.
We say that $S \PBn P$ iff $\qinit{S} \PBn \qinit{P}$.\footnotemark[\value{footnote}]
\end{defn}

\subsection{Controllability for nondeterministic automata}

Another lifting of controllability to the nondeterministic automata setting is, e.g., used by~\cite{flordal_supervision_2006} and \cite{flordal_compositional_2007}. We abbreviate this notion as $\FM$-controllability.

\begin{defn}[$\FM$-controllability]
\label{def:controllability-automata-flordal}
Let
$S = \aut{S}$ and $P = \aut{P}$ be nondeterministic automata.
A state $q_S \in \Q{S}$ is $\FM$-controllable \wrt{}\ state $q_P \in \Q{P}$, denoted $q_S \FM q_P$, iff
for all $w \in \Sigma^*$, $u \in \Sigmau$ and $q \in \Q{S}$, if $q_P \trans{wu}{P}$ and $q_S \trans{w}{S} q$, then also $q \trans{u}{S}$.
Supervisor $S$ is $\FM$-controllable \wrt{}\ $P$, denoted $S \FM P$, iff $\qinit{S} \FM \qinit{P}$.
\end{defn}

Note that this definition is slightly stronger than a direct automata-theoretic characterization of language controllability. The latter is obtained if we require that if $q_P \trans{wu}{P}$ and $q_S \trans{w}{S}$ then also $q_S \trans{wu}{S}$. That is, $\FM$-controllability requires that from the specific state $q$ such that $q_S \trans{w}{S} q$, it is possible to take the $u$-labelled transition, whereas this is not required by language controllability.

\subsection{$\Sigmau$-Admissibility}

Kushi and Takai~\cite{kushi_synthesis_2018}, define $\Sigmau$-admissibility.
Essentially a supervisor is $\Sigmau$-admissible if reachable states in the supervised plant do not disable uncontrollable actions. In the definition below $\mathit{Reach}(G) = \{ q \in Q \mid \exists_{w \in \Sigma^*}~q_0 \trans{w}{} q\}$ denotes the set of reachable states of $G$.

\begin{defn}[$\Sigmau$-Admissibility]\label{def:sigma-u-admissibility}
Let
$S = \aut{S}$ and $P = \aut{P}$ be nondeterministic automata. Then $S$ is $\Sigmau$-admissible \wrt\ $P$ if for all $(q_S,q_P) \in \mathit{Reach}(S \parallel P)$ and $u \in \Sigmau$, if $q_P \trans{u}{P}$ then $(q_S,q_P) \trans{u}{S\parallel P}$.
For the sake of brevity, we write $S \KT  P$ iff $S$ is $\Sigmau$-admissible \wrt\ $P$.
\end{defn}

\subsection{Nondeterministic State Controllability}

For a setting where plant, specification and supervisor are represented by nondeterministic automata, in \cite{zhou_control_2006}, state controllability is defined of a specification \wrt a plant. This definition is restricted to cases where the language of the specification is contained in the language of the plant. Also, the definition is partially formulated using languages instead of automata.

\begin{defn}[State controllability]
\label{def:state-controllability-zhou}
Let
$S = \aut{S}$ and $P = \aut{P}$ be nondeterministic automata such that $L(S) \subseteq L(P)$. Then, $S$ is state-controllable \wrt\ $P$ iff for all $w \in L(S)$ and $u \in \Sigmau$ such that $wu \in L(P)$, for all $q \in Q_S$, if $\qinit{S} \trans{w}{S} q$, then $q \trans{u}{S}$. 
For the sake of brevity, we write $S \SCn P$ iff $S$ is state-controllable (according to the above definition) \wrt\ $P$.
\end{defn}

Note that the restriction $L(S) \subseteq L(P)$ in the definition only has limited effect. If we drop the restriction, for $w \in L(S) \setminus L(P)$, the definition does not impose any restrictions since $w \notin L(P)$ implies $wu \notin L(P)$ for all $u \in \Sigmau$.
Kushi and Takai~\cite{kushi_synthesis_2018} show that $S$ is nondeterministic state controllable \wrt\ $P$ iff $S$ is $\Sigmau$-admissible \wrt\ $P$.

\subsection{Controllability for nondeterministic subautomata}

If the supervisor is a subautomaton of the plant and the supervisor can disable individual transitions, \cite{flordal_supervision_2006} and \cite{flordal_compositional_2007} use the following variation of $\FM$-controllability, that we abbreviate $\FMsubseteq$-controllability.
\begin{defn}[$\FMsubseteq$-Controllability]
\label{def:controllability-automata-flordal-subsupervisor}
Let
$S = \aut{S}$ and $P = \aut{P}$ be nondeterministic automata such that $S \subseteq P$.
A state $q_S \in \Q{S}$ is $\FMsubseteq$-controllable \wrt{}\ state $q_P \in \Q{P}$, denoted $q_S \FMsubseteq q_P$, iff for all $w \in \Sigma^*$, $u \in \Sigmau$, $q \in \Q{S}$ and $q' \in \Q{P}$, if $q_P \trans{w}{P} q \trans{u}{P} q'$ and $q_S \trans{w}{S} q$, then also $q \trans{u}{S} q'$.
Supervisor $S$ is $\FMsubseteq$-controllable \wrt{}\ $P$, denoted $S \FMsubseteq P$, iff $\qinit{S} \FMsubseteq \qinit{P}$.
\end{defn}

As mentioned by \cite{flordal_supervision_2006}, in case the supervisor is a subautomaton of a \emph{deterministic} plant, the notions of $\FM$-controllability and $\FMsubseteq$-controllability coincide.

\section{Comparing Notions of Controllability}
\label{sec:comparison}

We next compare the notions of controllability as they have been defined in the literature, for the contexts to which they can be applied. 
When $C_1$ and $C_2$ are controllability notions, we write  
$C_1 = C_2$ if $C_1$ and $C_2$ are equivalent.

In the remainder of this section, we first compare the notions of controllability that do not require the (language of the) supervisor to be a subautomaton of the plant. These cases are compared separately.
The notions are compared for each of the contexts for which they are defined.

\subsection{Nondeterministic Supervisors and Plants}

In this section, comparison of the previously defined notions of controllability is provided restricted to those notions that are defined for nondeterministic supervisors and plants. These are language controllability ($\LC$), nondeterministic partial bisimilarity ($\PBn$), FM-controllability ($\FM$) and $\Sigmau$-admissibility ($\KT$).

\begin{thm}\label{thm:nondet-S,nondet-P}
We have the following identifications and relations.
    \begin{center}
    \fbox
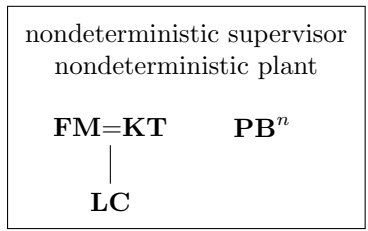
{\begin{tikzpicture}[node distance=1cm,align=center, uncontrollable/.style={densely dashed}]
\node (tekst) {\rm nondeterministic supervisor \\ \rm nondeterministic plant};
\node[below of=tekst,xshift=-1cm] (FMKT) {$\FM=\KT$};
\node[below of=FMKT] (LC) {$\LC$};
\node[right of=FMKT,xshift=1cm] (PBn) {$\PBn$};

\path (FMKT) edge node {} (LC);
\end{tikzpicture}}
\end{center}
\end{thm}

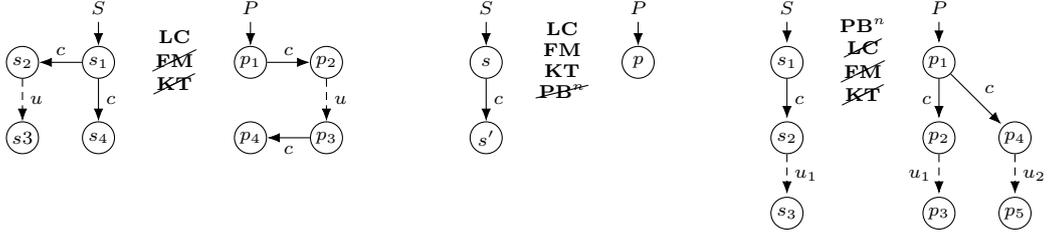
\begin{figure*}[htbp]
    \begin{center}
\begin{tikzpicture}[>=Latex,auto,node distance=1cm,align=center,font=\it, uncontrollable/.style={densely dashed},baseline={(0,0)}]
\scriptsize
\node[initial above, initial text = {$S$}, state] (s1) {$s_1$};
\node[state, left of=s1] (s2) {$s_2$};
\node[state, below of=s2] (s3) {$s3$};
\node[state, below of=s1] (s4) {$s_4$};
\node[right of = s1] (c) {$\begin{array}{c}\LC\\\cancel{\FM}\\\cancel{\KT}\end{array}$};
\node[initial above, initial text = {$P$},state, right of=c] (p1) {$p_1$};
\node[state, right of=p1] (p2) {$p_2$};
\node[state, below of=p2] (p3) {$p_3$};
\node[state, left of=p3] (p4) {$p_4$};
\path[->] (s1) edge node[above] {$c$} (s2);
\path[->] (s1) edge node {$c$} (s4);
\path[->] (s2) edge[uncontrollable] node {$u$} (s3);
\path[->] (p1) edge node {$c$} (p2);
\path[->] (p2) edge[uncontrollable] node {$u$} (p3);
\path[->] (p3) edge node {$c$} (p4);
\end{tikzpicture}
\qquad \qquad
\begin{tikzpicture}[>=Latex,auto,node distance=1cm,align=center,font=\it, uncontrollable/.style={densely dashed},baseline={(0,0)}]
\scriptsize
\node[initial above, initial text = {$S$}, state] (s) {$s$};
\node[state, below of=s] (s') {$s'$};
\node[right of = s] (c) {$\begin{array}{c}\LC\\\FM\\\KT\\\cancel{\PBn}\end{array}$};
\node[initial above, initial text = {$P$},state, right of=c] (p) {$p$};
\path[->] (s) edge node {$c$} (s');
\end{tikzpicture}
\qquad \qquad
\begin{tikzpicture}[>=Latex,shorten >=1pt,auto,->,node distance=1cm,align=center,font=\it, uncontrollable/.style={densely dashed},baseline={(0,0)}]
\scriptsize
\node[initial above, initial text = {$S$}, state] (s1) {$s_1$};
\node[state, below of=s1] (s2) {$s_2$};
\node[state, below of=s2] (s3) {$s_3$};

\path[->]
    (s1) edge node[right] {$c$} (s2)
    (s2) edge[uncontrollable] node[right] {$u_1$} (s3)
    ;

\node[right of = s1] (c) {$\begin{array}{c}\PBn\\\cancel{\LC}\\\cancel{\FM}\\\cancel{\KT}\end{array}$};

\node[initial above, initial text = {$P$}, state, right of=c] (p1) {$p_1$};
\node[state, below of=p1] (p2) {$p_2$};
\node[state, below of=p2] (p3) {$p_3$};
\node[state, below of=p1, right of=p1] (p4) {$p_4$};
\node[state, below of=p4] (p5) {$p_5$};

\path[->]
    (p1) edge node[left] {$c$} (p2)
    (p2) edge[uncontrollable] node[left] {$u_1$} (p3)
    (p1) edge node[above right] {$c$} (p4)
    (p4) edge[uncontrollable] node[right] {$u_2$} (p5)
    ;

\end{tikzpicture}
\end{center}
\caption{Examples showing absence of other relations in Theorem~\ref{thm:nondet-S,nondet-P}. 
}\label{fig:counterexamples-nondet-S-nondet-P}
\end{figure*}

We provide examples showing that additional implications cannot be added to this lattice in Figure~\ref{fig:counterexamples-nondet-S-nondet-P}.

\subsection{Deterministic Supervisors and Nondeterministic Plants}

In addition to the notions in the previous section, for deterministic supervisors and nondeterministic plants state controllability is also defined. 
Since this context is more specific than the context discussed in the previous subsection, all implications found there remain valid, but additional ones emerge as is shown in the following theorem.

\begin{thm}\label{thm:det-S-ndet-P}
We have the following identifications and relations.
\begin{center}
    \fbox{\begin{tikzpicture}[node distance=1cm,align=center, uncontrollable/.style={densely dashed}]
\node (tekst) {\rm deterministic supervisor\\ \rm nondeterministic plant};
\node[below of=tekst,xshift=-1cm] (ACFMKT) {$\LC=\FM=\KT$};
\node[right of=ACFMKT,xshift=1cm] (PBn) {$\PBn$};
\node[right of=PBn] (SC) {$\SC$};

\end{tikzpicture}}
\end{center}
\end{thm}

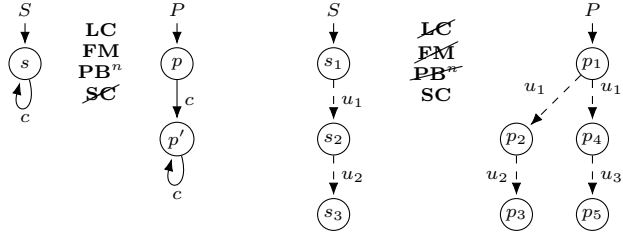
\begin{figure}[htbp]
\begin{center} 
\begin{tikzpicture}[>=Latex,shorten >=1pt,auto,->,node distance=1cm,align=center,font=\it, uncontrollable/.style={densely dashed},baseline={(0,0)}]
\scriptsize
\node[initial above, initial text = {$S$}, state] (s) {$s$};
\path[->] (s) edge[loop below] node[below] {$c$} (s);

\node[right of = s] (c) {$\begin{array}{c}\LC\\\FM\\\PBn\\\cancel{\SC}\end{array}$};

\node[initial above, initial text = {$P$}, state, right of=c] (p) {$p$};
\node[state, below of=p] (p') {$p'$};
\path[->] (p) edge node {$c$} (p');
\path[->] (p') edge[loop below] node[below] {$c$} (p');
\end{tikzpicture}
\qquad \qquad
\begin{tikzpicture}[>=Latex,shorten >=1pt,auto,->,node distance=1cm,align=center,font=\it, uncontrollable/.style={densely dashed},baseline={(0,0)}]
\scriptsize
\node[initial above, initial text = {$S$}, state] (s1) {$s_1$};
\node[state, below of=s1] (s2) {$s_2$};
\node[state, below of=s2] (s3) {$s_3$};

\path[->]
    (s1) edge[dashed] node[right] {$u_1$} (s2)
    (s2) edge[dashed] node[right] {$u_2$} (s3)
    ;

\node[right of = s1,xshift=10pt] (c) {$\begin{array}{c}\cancel{\LC}\\\cancel{\FM}\\\cancel{\PBn}\\\SC\end{array}$};

\node[initial above, initial text = {$P$}, state, right of=c, xshift=30pt] (p1) {$p_1$};
\node[state, below of=p1, left of=p1] (p2) {$p_2$};
\node[state, below of=p2] (p3) {$p_3$};
\node[state, below of=p1] (p4) {$p_4$};
\node[state, below of=p4] (p5) {$p_5$};

\path[->]
    (p1) edge[dashed] node[above left] {$u_1$} (p2)
    (p2) edge[dashed] node[left] {$u_2$} (p3)
    (p1) edge[dashed] node[above right] {$u_1$} (p4)
    (p4) edge[dashed] node[right] {$u_3$} (p5)
    ;
\end{tikzpicture}
\end{center}
\caption{Examples showing absence of other relations in Theorem~\ref{thm:det-S-ndet-P}.
}
\label{fig:counterexamples-det-S-nondet-P}
\end{figure}

Examples showing that additional implications cannot be added to this lattice are given in Figure~\ref{fig:counterexamples-det-S-nondet-P}. The right example in Figure~\ref{fig:counterexamples-nondet-S-nondet-P} shows that $\PBn$ does not imply the notions in the equivalence class of $\LC$ also in case the supervisor is deterministic.

\subsection{Deterministic Supervisors and Plants}

When we consider both supervisor and plant to be deterministic, two additional notions need to be compared. These are $\CR$ and $\PB$.

\begin{thm}
\label{thm:det-S,det-P}
We have the following identifications and relations.
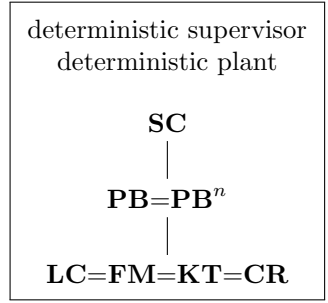
\begin{center}
    \fbox{\begin{tikzpicture}[node distance=1cm,align=center, uncontrollable/.style={densely dashed}]
\node (tekst) {\rm deterministic supervisor\\ \rm deterministic plant};
\node[below of=tekst] (SC) {$\SC$};
\node[below of=SC] (PB) {$\PB=\PBn$};
\node[below of=PB] (LCACCRFM) {$\LC=\FM=\KT=\CR$};

\path (SC) edge node {} (PB);
\path (PB) edge node {} (LCACCRFM);
\end{tikzpicture}}
\end{center}
\end{thm}

No additional implications can be added to the lattice in Theorem~\ref{thm:det-S,det-P}.
The left example in Figure~\ref{fig:counterexamples-det-S-nondet-P} shows that $\PB$ and $\PBn$ do not imply $\SC$. The middle example in Figure~\ref{fig:counterexamples-nondet-S-nondet-P} shows that the notions in the equivalence class of $\LC$ do not imply $\PBn$ and $\PB$.

\subsection{Subautomata}

Many implementations deciding controllability use the premise that the supervisor is a subautomaton of the plant, that is $S \subseteq P$.
In this case, additionally $\FMsubseteq$ and $\SCn$ are defined (as $S \subseteq P$ implies $L(S) \subseteq L(P)$).
Both notions are defined for the general context of nondeterministic supervisor and plant. For the sake of brevity we restrict our exposition to that context.

\begin{thm}\label{thm:unlabeled}
We have the following identifications and relations.

    \begin{center}
    \fbox
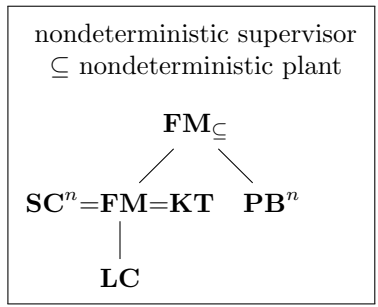
{\begin{tikzpicture}[node distance=1cm,align=center, uncontrollable/.style={densely dashed}]
\node (tekst) {\rm nondeterministic supervisor\\$\subseteq$ \rm nondeterministic plant};
\node[below of=tekst] (FMsubseteq) {$\FMsubseteq$};
\node[below of=FMsubseteq, left of=FMsubseteq] (SCnFMKT) {$\SCn = \FM=\KT$};
\node[below of=SCnFMKT] (AC) {$\LC$};
\node[below of=FMsubseteq, right of=FMsubseteq] (PBn) {$\PBn$};

\path (FMsubseteq) edge (SCnFMKT)
      (FMsubseteq) edge (PBn)
      (SCnFMKT) edge node {} (AC);
\end{tikzpicture}}
\end{center}
\end{thm}

\begin{figure*}[htbp]
\begin{center} 
\begin{tikzpicture}[>=Latex,shorten >=1pt,auto,->,node distance=1cm,align=center,font=\it, uncontrollable/.style={densely dashed},baseline={(0,0)}]
\scriptsize
\node[initial above, initial text = {$S$}, state] (s1) {$p_1$};
\node[state, below of=s1] (s2) {$p_2$};
\node[state, below of=s2, left of = s2] (s3) {$p_3$};

\path[->]
    (s1) edge node[left] {$c$} (s2)
    (s2) edge[uncontrollable] node[above left] {$u$} (s3)
    ;

\node[right of = s1] (c) {$\begin{array}{c}\cancel{\FMsubseteq}\\\FM\\\PBn\end{array}$};

\node[initial above, initial text = {$P$}, state, right of=c] (p1) {$p_1$};
\node[state, below of=p1] (p2) {$p_2$};
\node[state, below of=p2, left of=p2] (p3) {$p_3$};
\node[state, below of=p2, right of=p2] (p4) {$p_4$};

\path[->]
    (p1) edge node[left] {$c$} (p2)
    (p2) edge[uncontrollable] node[above left] {$u$} (p3)
    (p2) edge[uncontrollable] node[above right] {$u$} (p4)
    ;
\end{tikzpicture}
\qquad \qquad
\begin{tikzpicture}[>=Latex,shorten >=1pt,auto,->,node distance=1cm,align=center,font=\it, uncontrollable/.style={densely dashed},baseline={(0,0)}]
\scriptsize
\node[initial above, initial text = {$S$}, state] (s1) {$p_1$};
\node[state, below of=s1, left of=s1] (s2) {$p_2$};
\node[state, below of=s1] (s3) {$p_3$};
\node[state, below of=s3] (s4) {$p_4$};

\path[->]
    (s1) edge node[above left] {$c$} (s2)
    (s1) edge node[right] {$c$} (s3)
    (s3) edge[uncontrollable] node[right] {$u$} (s4)
    ;

\node[right of = s1] (c) {$\begin{array}{c}\LC\\\cancel{\FM}\end{array}$};

\node[initial above, initial text = {$P$}, state, right of=c] (p1) {$p_1$};
\node[state, below of=p1] (p2) {$p_2$};
\node[state, below of=p1, right of=p1] (p3) {$p_3$};
\node[state, below of=p3] (p4) {$p_4$};

\path[->]
    (p1) edge node[left] {$c$} (p2)
    (p1) edge node[above right] {$c$} (p3)
    (p3) edge[uncontrollable] node[right] {$u$} (p4)
    ;
\end{tikzpicture}
\qquad \qquad
\begin{tikzpicture}[>=Latex,shorten >=1pt,auto,->,node distance=1cm,align=center,font=\it, uncontrollable/.style={densely dashed},baseline={(0,0)}]
\scriptsize
\node[initial above, initial text = {$S$},state] (s1) {$p_1$};
\node[state, below of=s1] (s2) {$p_2$};
\node[state, below of=s2] (s4) {$p_4$};

\path[->]
    (s1) edge[uncontrollable] node[right] {$u$} (s2)
    (s2) edge node[right] {$c$} (s4)
    ;

\node[right of = s1] (c) {$\begin{array}{c}\cancel{\PBn}\\
\LC\\
\FM
\end{array}$};

\node[initial above, initial text = {$P$},state, right of=c] (p1) {$p_1$};
\node[state, below of=p1] (p2) {$p_2$};
\node[state, below of=p1, right of=p1] (p3) {$p_3$};
\node[state, below of=p2] (p4)  {$p_4$};

\path[->]
    (p1) edge[uncontrollable] node[above left] {$u$} (p2)
    (p1) edge[uncontrollable] node[above right] {$u$} (p3)
    (p2) edge  node[right]  {$c$} (p4)
    ;
\end{tikzpicture}
\end{center}
\caption{Examples showing absence of other relations in Theorem~\ref{thm:unlabeled}.
\label{fig:counterexamples-nondet-S-nondet-P-subset}
}
\end{figure*}
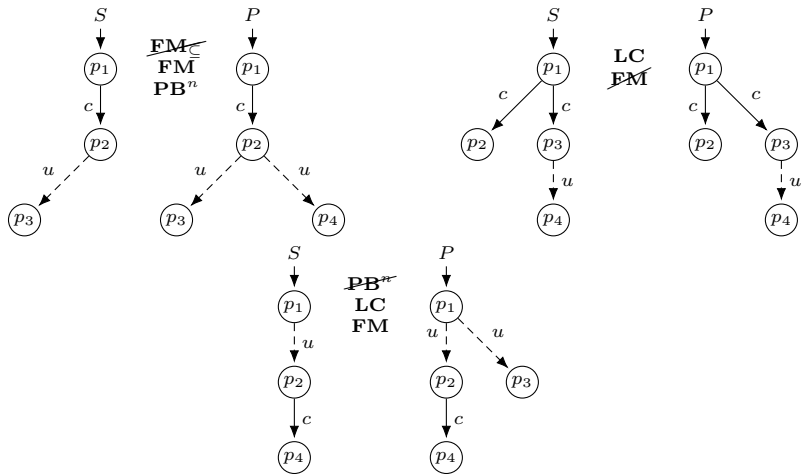

Examples showing that additional implications cannot be added to the lattice in Theorem~\ref{thm:unlabeled} are given in Figure~\ref{fig:counterexamples-nondet-S-nondet-P-subset}.

\section{Discussion}
\label{sec:discussion}

\newcommand{\X}{\mathrel{\textbf{X}}}

In the previous section we have seen that, depending on the context, very different relationships between the notions emerge. We would insist that, in any context, a notion of controllability should at least imply language controllability.
When the plant is nondeterministic, only $\LC$, $\FM$ and $\KT$ satisfy this requirement, regardless of whether the supervisor is deterministic or not.
When plant and supervisor are both deterministic all controllability notions imply language controllability.

From a practical point of view, it is necessary to have controllability of the supervised plant \wrt\ the plant, instead of the supervisor \wrt\ the plant. 
Since the supervised plant $S \parallel P$ is more closely related to plant $P$ than an arbitrary supervisor $S$, more relationships may result.
In the remainder of this section, we therefore reconsider all defined notions of controllability, where instead of the supervisor $S$ we consider the supervised plant $S \parallel P$.  

If $S$ and $P$ are deterministic, then $S \parallel P$ is deterministic and $S \parallel P \subseteq S$ and $S \parallel P \subseteq P$. Therefore, in this setting also $\SCn$ and $\FMsubseteq$ are defined, and both notions imply language controllability according to Theorem~\ref{thm:unlabeled}. Hence, in this setting all notions of controllability identified in this paper are defined and imply language controllability.

If $S$ or $P$ is nondeterministic, then $S \parallel P$ 
may also be nondeterministic, so we only consider the setting where both automata (supervised plant and plant) are nondeterministic. 
In this case we have $L(S \parallel P) \subseteq L(P)$ but not necessarily $S \parallel P \subseteq P$. So, in this setting also $\SCn$ is defined, but $\FMsubseteq$ is not. According to Theorem~\ref{thm:unlabeled}, $\SCn$, $\FM$, $\KT$ and $\LC$ imply language controllability. The following example illustrates that $\PBn$ does not imply $\LC$.

\begin{exmp}
\label{exa13}
Consider the following supervisor $S$ and plant $P$. Note that $(S \parallel P) \PBn P$.
However, it is not the case that $(S \parallel P) \LC P$.
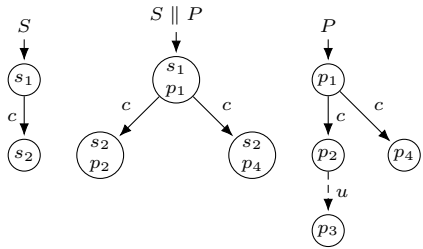
\begin{center} 
\begin{tikzpicture}[>=Latex,shorten >=1pt,auto,->,node distance=1cm,align=center,font=\it, uncontrollable/.style={densely dashed}]
\scriptsize
\node[initial above, initial text = {$S$}, state] (s1) {$s_1$};
\node[state, below of=s1] (s2) {$s_2$};

\path[->]
    (s1) edge node[left] {$c$} (s2)
    ;

\node[state, right of=s2] (sp2) {$s_2$\\$p_2$};
\node[initial above, initial text = {$S \parallel P$}, state, above of=sp2, right of = sp2] (sp1) {$s_1$\\$p_1$};

\node[state, below of=sp1, right of=sp1] (sp4) {$s_2$\\$p_4$};

\path[->]
    (sp1) edge node[above left] {$c$} (sp2)
    (sp1) edge node[above right] {$c$} (sp4)
    ;
    
\node[state, right of=sp4] (p2) {$p_2$};
\node[initial above, initial text = {$P$}, state, above of=p2] (p1) {$p_1$};
\node[state, below of=p2] (p3) {$p_3$};
\node[state,right of=p2] (p4) {$p_4$};

\path[->]
    (p1) edge node[right] {$c$} (p2)
    (p2) edge[uncontrollable] node[right] {$u$} (p3)
    (p1) edge node[above right] {$c$} (p4)
    ;

\end{tikzpicture}
\end{center}
\end{exmp}

\section{Conclusion}
\label{sec:conclusion}

In this paper, several notions of controllability of a supervisor w.r.t. a plant have been compared in different contexts and relations between these notions are mathematically characterized.

In the general context where supervisor and plant are nondeterministic, the equivalent notions of nondeterministic controllability by Flordal and Malik ($\FM$-controllability)~\cite{flordal_supervision_2006,flordal_compositional_2007} and $\Sigmau$-admissibility by Kushi and Takai ($\KT$-controllability)~\cite{kushi_synthesis_2018} are the only notions that imply language controllability.

From a practical perspective, it is more interesting to consider the controllability of the supervised plant \wrt the plant.  
We conclude that the notions of $\FM$-controllability, $\Sigmau$-admissibility, and (nondeterministic) state controllability~\cite{zhou_control_2006} are equally appropriate for use when considering controllability of the supervised plant \wrt\ the plant.

So, when considering all contexts discussed in this paper, only $\FM$-controllability and $\Sigmau$-admissibility always imply language controllability.

The notion of observability 
is of similar importance in the field of supervisory control theory as controllability. Like controllability, it is used in different contexts, with different formalizations, and unexplored relationships between those.
As future work, observability could be classified in a similar way. 

\bibliographystyle{plain}
\bibliography{References}

\appendix
\section{Proofs}
\label{sec:proofs}

In the appendix, we use the following additional notation.

Given a state $q \in Q$ and $w \in \Sigma^*$, we write $\Delta(q,w) = \{ q' \in Q \mid q \trans{w}{} q' \}$. 
If $G$ is deterministic and $w \in L(G)$, $\Delta(\qinit{}, w)$ is a singleton, and with a slight abuse of notation we write $\Delta(\qinit{}, w)$ also for the single state that it contains.

We also use the following notation. Let $C_1$ and $C_2$ be two controllability notions, we write $C_1 \subseteq C_2$ if $C_1$ is stronger than $C_2$, $C_1 = C_2$ if $C_1$ and $C_2$ are equivalent, i.e., $C_1 \subseteq C_2$ and $C_2 \subseteq C_1$, $C_1 \subsetneq C_2$ if $C_1 \subseteq C_2$ and not $C_1 = C_2$. We write $C_1 \incomparable C_2$ if $C_1$ and $C_2$ are incomparable, that is, neither $C_1 \subseteq C_2$ nor $C_2 \subseteq C_1$. 

\medskip
\noindent
Recall that Lemma~\ref{lem:cc} shows that $L(S) \LC L(P)$ iff $L(S \parallel P) \LC L(P)$.

\noindent
{\it Proof of Lemma~\ref{lem:cc}.}
Trivial. \qed
\medskip

To facilitate the easier comparison between the notions of controllability to be introduced in the remainder of this paper, we first reformulate language controllability on the states of automata. 

\begin{defn}[Automata controllability]
\label{def:controllability-automata}
Consider nondeterministic supervisor $S = \aut{S}$ and nondeterministic plant $P= \aut{P}$.
A state $q_S \in \Q{S}$ is automata controllable \wrt state $q_P \in \Q{P}$, denoted $q_S \AC q_P$, iff
for all $w \in \Sigma^*$, $u \in \Sigmau$ if $q_P \trans{wu}{P}$ and $q_S \trans{w}{S} $, then also $q_S \trans{wu}{S}$.
Supervisor $S$ is automata controllable \wrt $P$, denoted $S \AC P$ iff $\qinit{S} \AC \qinit{P}$.
\end{defn}

The notion of automata controllability and language controllability agree as expressed by the following lemma.

\begin{lem}[$\AC=\LC$]\label{lem:AC-vs-LC}
Let $S$ and $P$ be nondeterministic automata. Then $S \AC P$ iff $L(S) \LC L(P)$.
\end{lem}

\begin{proof}{~} 
\begin{description}
\item[$\Rightarrow$] Suppose that $S \AC P$. We prove for all $w \in \Sigma^*$, $u \in \Sigmau$, if $wu \in L(S)\Sigmau \cap L(P)$, then $wu \in L(S)$.
Pick arbitrary $wu \in L(S)\Sigmau \cap L(P)$. Then, $wu \in L(P)$, so $\qinit{P} \trans{wu}{P}$, and $w \in L(S)$, so $\qinit{S} \trans{w}{S}$.
As $S \AC P$, $\qinit{S}\trans{wu}{S}$, so $wu \in L(S)$, hence $L(S) \LC L(P)$. 

\item[$\Leftarrow$]Suppose that $L(S) \LC L(P)$. We prove for all $w \in \Sigma^*$, $u \in \Sigmau$, if $\qinit{P} \trans{wu}{P}$ and $\qinit{S} \trans{w}{S}$ then $\qinit{S} \trans{wu}{S}$.
Pick arbitrary $w \in \Sigma^*$, $u \in \Sigmau$ such that $\qinit{P} \trans{wu}{P}$ and $\qinit{S} \trans{w}{S}$, and note that $wu \in L(P)$ and $w \in L(S)$, so $wu \in L(S)\Sigmau\cap L(P)$. As $L(S) \LC L(P)$, $wu \in L(S)$, so $\qinit{S} \trans{wu}{S}$. Hence $S \AC P$. \qedhere
\end{description}
\end{proof}
In view of the previous lemma, we refrain from considering language controllability in the comparisons to come, and refer to automata controllability instead.

\subsection{Nondeterministic Supervisors and Plants}

In this section, we provide the details required to establish the results in Theorem~\ref{thm:nondet-S,nondet-P}, capturing the relations between controllability notions for nondeterministic supervisors and plants.

\begin{lem}[$\FM=\KT$]
\label{lem:FM-is-KT}
For nondeterministic automata $S$ and $P$: $S \FM P$ iff $S \KT P$.
\end{lem}

\begin{proof}{~} 
    \begin{description}
    \item[$\Rightarrow$] Suppose that $S  \FM P$. Let $(q_S,q_P) \in \mathit{Reach}(S \parallel P)$ and let $u \in \Sigmau$. Assume that $q_P \trans{u}{P}$. We need to show that $(q_S,q_P) \trans{u}{S \parallel P}$. Since $(q_S,q_P) \in \mathit{Reach}(S \parallel P)$, we have the existence of $w \in \Sigma^*$ such that $(\qinit{S},\qinit{P}) \trans{w}{S \parallel P} (q_S,q_P)$. Therefore, as $S$ and $P$ have the same event set and by definition of $\parallel$, we obtain $\qinit{S} \trans{w}{S} q_S$ and $\qinit{P} \trans{w}{P} q_P$. Combining $\qinit{P} \trans{w}{P} q_P$ and $q_P \trans{u}{P}$ gives $\qinit{P} \trans{wu}{P}$. Since $S$ and $P$ are FM-controllable, combined with $\qinit{P} \trans{wu}{P}$ and $\qinit{S} \trans{w}{S} q_S$, it follows that $q_S \trans{u}{S}$. Thus we have both $q_S \trans{u}{S}$ and $q_P \trans{u}{P}$, and thus $(q_S,q_P) \trans{u}{S \parallel P}$ as well, which was to be proven.
    \item[$\Leftarrow$] Suppose that $S \KT P$. Let $w \in \Sigma^*$, $u \in \Sigmau$ and $q_S \in Q_S$. Assume that $\qinit{P} \trans{wu}{P}$ and $\qinit{S} 
    \trans{w}{S} q_S$ We need to show that $q_S \trans{u}{S}$.
    From $\qinit{P} \trans{wu}{P}$ we have the existence of $q_P \in Q_P$ such that $\qinit{P} \trans{w}{P} q_P$ and $q_P \trans{u}{P}$.
    Since $S$ and $P$ have the same event set, and by definition of $\parallel$ it follows that $(\qinit{S},\qinit{P}) \trans{w}{S \parallel P} (q_S,q_P)$. Since $(q_S,q_P) \in \mathit{Reach}(S \parallel P)$ and $q_P \trans{u}{P}$ and because $S$ is $\Sigmau$-admissible \wrt\ $P$, we have $(q_S,q_P) \trans{u}{S \parallel P}$. Thus, also $q_S \trans{u}{S}$, which was to be shown. \qedhere
\end{description}
\end{proof}

\begin{lem}[$\FM \subseteq \AC$]\label{lem:FM-implies-AC}
For nondeterministic automata $S$ and $P$: If $S \FM P$, then $S \AC P$.
\end{lem}

\begin{proof} 
Let $S = \aut{S}$ and $P = \aut{P}$ be nondeterministic automata.
Assume $S \FM P$. 
Let $w \in \Sigma^*$, $u \in \Sigmau$. Assume $\qinit{p} \trans{wu}{P}$ and $\qinit{S} \trans{w}{S}$. We need to prove that $\qinit{S} \trans{wu}{S}$.
From $\qinit{S} \trans{w}{S} $ we have $\qinit{S} \trans{w}{S} q_S$ for some $q_S \in Q_S$. 
As $S \FM P$, we have $q_S \trans{u}{S}$, and consequently  $\qinit{S} \trans{wu}{S}$. \qedhere
\end{proof}

%Since we have already established that $\FM = \KT$, we forego the discussion of $\Sigmau$-admissibility ($\KT$) in the following lemmas in this section, and only consider $\FM$-controllability.

\medskip
\noindent
{\it Proof of Theorem~\ref{thm:nondet-S,nondet-P}.}
    \begin{itemize}
        \item $\FM=\KT$ follows from Lemma~\ref{lem:FM-is-KT};
        \item $\FM \subseteq \AC$ and $\KT \subseteq \AC$ follow from Lemma~\ref{lem:FM-implies-AC} and $\FM=\KT$;
        \item $\AC \nsubseteq \FM$ follows from Example~\ref{exa:nondet-S-nondet-P-AC-does-not-imply-FM};
        \item $\PBn \incomparable \FM$, $\PBn \incomparable \KT$ and $\PBn \incomparable \AC$ follow from Example~\ref{exa:if-det-S,det-P-then-AC,FM-does-not-imply-PB} and Example~\ref{exa:if-det-S-nondet-P-then-PB-does-not-imply-FM-AC}.  \qed
    \end{itemize}

\begin{exmp}\label{exa:nondet-S-nondet-P-AC-does-not-imply-FM}
Consider the following nondeterministic supervisor $S$ and (determinstic) plant $P$.

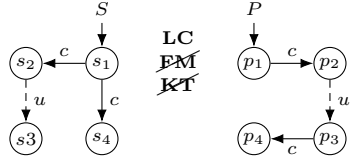
\begin{center}
\begin{tikzpicture}[>=Latex,auto,node distance=1cm,align=center,font=\it, uncontrollable/.style={densely dashed}]
\scriptsize
\node[initial above, initial text = {$S$}, state] (s1) {$s_1$};
\node[state, left of=s1] (s2) {$s_2$};
\node[state, below of=s2] (s3) {$s3$};
\node[state, below of=s1] (s4) {$s_4$};
\node[right of = s1] (c) {$\begin{array}{c}\LC\\\cancel{\FM}\\\cancel{\KT}\end{array}$};
\node[initial above, initial text = {$P$},state, right of=c] (p1) {$p_1$};
\node[state, right of=p1] (p2) {$p_2$};
\node[state, below of=p2] (p3) {$p_3$};
\node[state, left of=p3] (p4) {$p_4$};
\path[->] (s1) edge node[above] {$c$} (s2);
\path[->] (s1) edge node {$c$} (s4);
\path[->] (s2) edge[uncontrollable] node {$u$} (s3);
\path[->] (p1) edge node {$c$} (p2);
\path[->] (p2) edge[uncontrollable] node {$u$} (p3);
\path[->] (p3) edge node {$c$} (p4);
\end{tikzpicture}
\end{center}

We have:
\begin{itemize}
    \item $S \LC P$;
    \item $S \mathrel{\cancel{\FM}} P$ as $p_1 \trans{cu}{}$, and $s_1 \trans{c}{} s_4$, but $s_4 \ntrans{u}{}$;
    \item $S \mathrel{\cancel{\KT}} P$ as $\FM = \KT$.
\end{itemize}
\end{exmp}

\begin{exmp}
\label{exa:if-det-S,det-P-then-AC,FM-does-not-imply-PB}
Consider the following supervisor $S$ and plant $P$, both with event set $\Sigma = \{c\}$.
\begin{center}
\begin{tikzpicture}[>=Latex,auto,node distance=1cm,align=center,font=\it, uncontrollable/.style={densely dashed}]
\scriptsize
\node[initial above, initial text = {$S$}, state] (s) {$s$};
\node[state, below of=s] (s') {$s'$};
\node[right of = s] (c) {$\begin{array}{c}\LC\\\FM\\\KT\\\cancel{\PBn}\end{array}$};
\node[initial above, initial text = {$P$},state, right of=c] (p) {$p$};
\path[->] (s) edge node {$c$} (s');
\end{tikzpicture}
\end{center}
We have $S \LC P$, $S \FM P$, and $S \KT P$, but not $S \PBn P$ because no partial bisimulation relation between $s$ and $p$ exists as $p$ cannot mimic the $c$-transition.
So, $\LC \not \subseteq \PBn$, $\FM \not \subseteq \PBn$, and $\KT \not\subseteq \PBn$.
\end{exmp}

\begin{exmp}
\label{exa:if-det-S-nondet-P-then-PB-does-not-imply-FM-AC}
Consider the following deterministic supervisor $S$ and nondeterministic plant $P$, with $\Sigmau = \{u_1, u_2\}$.
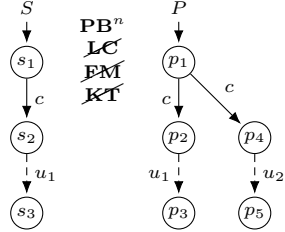
\begin{center} 
\begin{tikzpicture}[>=Latex,shorten >=1pt,auto,->,node distance=1cm,align=center,font=\it, uncontrollable/.style={densely dashed}]
\scriptsize
\node[initial above, initial text = {$S$}, state] (s1) {$s_1$};
\node[state, below of=s1] (s2) {$s_2$};
\node[state, below of=s2] (s3) {$s_3$};

\path[->]
    (s1) edge node[right] {$c$} (s2)
    (s2) edge[uncontrollable] node[right] {$u_1$} (s3)
    ;

\node[right of = s1] (c) {$\begin{array}{c}\PBn\\\cancel{\LC}\\\cancel{\FM}\\\cancel{\KT}\end{array}$};

\node[initial above, initial text = {$P$}, state, right of=c] (p1) {$p_1$};
\node[state, below of=p1] (p2) {$p_2$};
\node[state, below of=p2] (p3) {$p_3$};
\node[state, below of=p1, right of=p1] (p4) {$p_4$};
\node[state, below of=p4] (p5) {$p_5$};

\path[->]
    (p1) edge node[left] {$c$} (p2)
    (p2) edge[uncontrollable] node[left] {$u_1$} (p3)
    (p1) edge node[above right] {$c$} (p4)
    (p4) edge[uncontrollable] node[right] {$u_2$} (p5)
    ;

\end{tikzpicture}
\end{center}
We have that $R = \{ (s_1, p_1), (s_2, p_2), (s_3, p_3) \}$ is a partial bisimulation relation.
However, $S$ is not language controllable w.r.t. $P$, since $cu_2 \in L(P)$, $c \in L(S)$, but $cu_2 \not\in L(S)$.
The argumentations for $S \mathrel{\cancel{\FM}} P$ and $S \mathrel{\cancel{\KT}} P$ are similar.
\end{exmp}

\subsection{Deterministic Supervisors and Nondeterministic Plants}

In this section, we provide the details required to establish the results in Theorem~\ref{thm:det-S-ndet-P}, capturing the relations between controllability notions for deterministic supervisors and nondeterministic plants.

In this context, automata controllability is stronger than FM-controllability.
\begin{lem}[$\AC\subseteq\FM$]
\label{lem:controllable-implies-flordal-malik-controllable}
Let $S$ be a deterministic automaton and $P$ a (possibly nondeterministic) automaton. If $S \AC P$, then $S \FM P$.
\end{lem}
\begin{proof} 
Let $S = \aut{S}$ be a deterministic automaton and $P = \aut{P}$ a nondeterministic automaton.
Assume that $S \AC P$. We show that $S \FM P$.
Fix $w \in \Sigma^*$, $u \in \Sigmau$ and $q \in Q_S$ such that $\qinit{P} \trans{wu}{P}$ and $\qinit{S} \trans{w}{S} q$. 
Since $S \AC P$, $\qinit{S} \trans{wu}{S}$; since $S$ is deterministic, $q =\Delta(\qinit{S}, w)$ is unique, so $q \trans{u}{S}$. Therefore, $S \FM P$.
\end{proof}

\noindent
{\it Proof of Theorem~\ref{thm:det-S-ndet-P}.}
  \begin{itemize}
      \item $\FM = \KT$ follows from Lemma~\ref{lem:FM-is-KT}; we do not consider $\KT$ explicitly in the proofs below.
      \item $\AC = \FM$ follows from Lemmas~\ref{lem:FM-implies-AC} and~\ref{lem:controllable-implies-flordal-malik-controllable};
      \item $\PBn \incomparable \AC$ and $\PBn \incomparable \FM$ follow from Examples~\ref{exa:if-det-S,det-P-then-AC,FM-does-not-imply-PB} and~\ref{exa:if-det-S-nondet-P-then-PB-does-not-imply-FM-AC}, which remain valid since $S$ is deterministic;
      \item $\SC \incomparable \AC$, $\SC \incomparable \FM$, and $\SC \incomparable \PBn$ follow from Examples~\ref{exa:if-det-S,det-P-then-AC,FM,PB-does-not-imply-SC} and~\ref{exa:if-det-S-nondet-P-then-SC-does-not-imply-FM-AC-PB}. \qedhere
  \end{itemize}

\begin{exmp}
\label{exa:if-det-S,det-P-then-AC,FM,PB-does-not-imply-SC}
Consider the following deterministic supervisor $S$ and plant $P$.
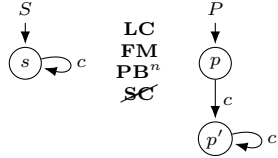
\begin{center} 
\begin{tikzpicture}[>=Latex,shorten >=1pt,auto,->,node distance=1cm,align=center,font=\it, uncontrollable/.style={densely dashed}]
\scriptsize
\node[initial above, initial text = {$S$}, state] (s) {$s$};
\path[->] (s) edge[loop right] node[right] {$c$} (s);

\node[right of = s,xshift=0.5cm] (c) {$\begin{array}{c}\LC\\\FM\\\PBn\\\cancel{\SC}\end{array}$};

\node[initial above, initial text = {$P$}, state, right of=c] (p) {$p$};
\node[state, below of=p] (p') {$p'$};
\path[->] (p) edge node {$c$} (p');
\path[->] (p') edge[loop right] node[right] {$c$} (p');
\end{tikzpicture}
\end{center}
Note that $R = \{ (s,p), (s,p') \}$ is a nondeterministic partial bisimulation relation. However, there exists no embedding $f$ of $S$ to $P$ witnessing $S \SC P$, so 
$\PBn \not \subseteq \SC$. Both $S \LC P$ and $S \FM P$ trivially hold since the plant and supervisor do not have any uncontrollable events.
\end{exmp}

\begin{exmp}
\label{exa:if-det-S-nondet-P-then-SC-does-not-imply-FM-AC-PB}
Consider the following deterministic supervisor $S$ and nondeterministic plant $P$, with $\Sigmau = \{u_1,u_2,u_3\}$.
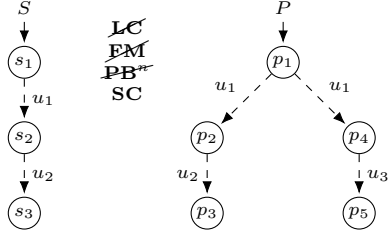
\begin{center} 
\begin{tikzpicture}[>=Latex,shorten >=1pt,auto,->,node distance=1cm,align=center,font=\it, uncontrollable/.style={densely dashed}]
\scriptsize
\node[initial above, initial text = {$S$}, state] (s1) {$s_1$};
\node[state, below of=s1] (s2) {$s_2$};
\node[state, below of=s2] (s3) {$s_3$};

\path[->]
    (s1) edge[dashed] node[right] {$u_1$} (s2)
    (s2) edge[dashed] node[right] {$u_2$} (s3)
    ;

\node[right of = s1,xshift=10pt] (c) {$\begin{array}{c}\cancel{\LC}\\\cancel{\FM}\\\cancel{\PBn}\\\SC\end{array}$};

\node[initial above, initial text = {$P$}, state, right of=c, xshift=30pt] (p1) {$p_1$};
\node[state, below of=p1, left of=p1] (p2) {$p_2$};
\node[state, below of=p2] (p3) {$p_3$};
\node[state, below of=p1, right of=p1] (p4) {$p_4$};
\node[state, below of=p4] (p5) {$p_5$};

\path[->]
    (p1) edge[dashed] node[above left] {$u_1$} (p2)
    (p2) edge[dashed] node[left] {$u_2$} (p3)
    (p1) edge[dashed] node[above right] {$u_1$} (p4)
    (p4) edge[dashed] node[right] {$u_3$} (p5)
    ;

\end{tikzpicture}
\end{center}

We have that $S \SC P$, witnessed by $f(s_1) = p_1, f(s_2) = p_2$ and $f(s_3) = p_3$. In particular, no uncontrollable events are forbidden in $p_1$, even though the transition to $p_4$ is removed. 
However, $S$ is not language controllable w.r.t. $P$, since $u_1u_3 \in L(P)$, $u_1 \in L(S)$, but  $u_1u_3 \notin L(S)$. The argumentation for $S \mathrel{\cancel{\FM}} P$ is similar.
Also, there is no partial bisimulation relation relating $s_2$ in the supervisor and $p_4$ in the plant, since $s_2$ cannot mimic the transition to $p_5$, however, in any partial bisimulation relation, $s_2$ and $p_4$ must be related.
\end{exmp}

\subsection{Deterministic Supervisors and Plants}

In this section, we provide the details required to establish the results in Theorem~\ref{thm:det-S,det-P}, capturing the relations between controllability notions for deterministic supervisors and plants.

We first establish that for deterministic supervisors and plants, automata controllability and control relations coincide.
This result was originally established by \cite{rutten_coalgebra_2000} in the setting of language controllability. 
\begin{lem}[$\AC = \CR$]
\label{lem:if-det-S-det-P-then-AC-is-CR}
Let $S$ and $P$ be deterministic automata.
Then $S \AC P$ iff $S \CR P$.
\end{lem}

\begin{proof} 
Let $S =\aut{S}$ and $P = \aut{P}$ be deterministic automata.
\begin{itemize}
    \item[$\Rightarrow$]
    Suppose $S \AC P$. Let $R = \{ (q_S, q_P) \mid  q_S \AC q_P \}$. We check that $R$ is a control relation. Pick arbitrary $q_S, q_P$ such that $q_S \mathrel{R} q_P$.
    We prove both conditions on control relations:
    \begin{itemize}
        \item Let $a \in \Sigma$ such that $q_S \trans{a}{S} q'_S$ and $q_P \trans{a}{P} q'_P$. Towards a  contradiction, suppose $(q'_S,q'_P) \notin R$, then  $q'_S$ is not automata controllable \wrt $q'_P$, so there exist $w \in \Sigma^*$, $u \in \Sigmau$ such that $q'_P \trans{wu}{P}$ and $q'_S \trans{w}{S}$, but $q'_S \ntrans{wu}{S}$. Then $aw \in \Sigma^*$, $u \in \Sigmau$, and  $q_P \trans{awu}{P}$ and $q_S \trans{aw}{S}$.
        Because $S$ and $P$ are deterministic, $\Delta(q'_P,wu) = \Delta(q_P,awu)$ and $\Delta(q'_S,w) = \Delta(q_S,aw)$, hence $q_S \ntrans{awu}{S}$, but then not $q_S \AC q_P$, which is a  contradiction.
    
        \item Let $u  \in \Sigmau$, $q'_P$ such that $q_P \trans{u}{P} q'_P$. As $q_S \AC q_P$, $q_S \trans{u}{S} q'_S$, for some (unique) $q'_S$. We  must show $q'_S \AC q'_P$.  The argument is analogous to the argument in the previous case.
    \end{itemize}

    \item[$\Leftarrow$] 
    Suppose $S \CR P$. Then there exists a control relation $R$ such that $\qinit{S} \R \qinit{P}$. We prove that $\qinit{S} \AC \qinit{P}$.
We need to prove for all $w \in \Sigma^*$, $u \in \Sigmau$, if $\qinit{P} \trans{wu}{P}$ and $\qinit{S} \trans{w}{S}$ that $\qinit{S} \trans{wu}{S}$.

To do so, we establish the following result.
For all $w \in \Sigma^*$, $u \in \Sigmau$, and for all $q_S, q_P$ such that $q_S \R q_P$, if $q_P \trans{wu}{P}$ and $q_S \trans{w}{S}$, then $q_S \trans{wu}{S}$.
We proceed by induction on $w$.
\begin{itemize}
    \item $w = \epsilon$. Fix arbitrary $q_S, q_P$ such that $q_S \R q_P$, and suppose $q_P \trans{u}{P}$. As $q_S \R q_P$, there exists some $q'_S$ such that $q_S \trans{u}{S} q'_S$.
    \item $w = aw'$. Fix arbitrary $q_S, q_P$ such that $q_S \R q_P$, and suppose $q_P \trans{aw'u}{P}$ and $q_S \trans{aw'}{S}$. So, there exist $q'_P$, $q'_S$ such that $q_P \trans{a}{P} q'_P \trans{w'u}{P}$, and $q_S \trans{a}{S} q'_S \trans{w'}{S}$. As $q_P \R q_S$, it must be the case that $q'_S \R q'_P$ by definition of control relation.
    According to the induction hypothesis $q'_S \trans{w'u}{S}$.
    It follows immediately that also $q_S \trans{aw}{S}$.
\end{itemize}
As $\qinit{S} \R \qinit{P}$ the desired result follows. \qedhere
\end{itemize}
\end{proof}

Also, the different notions of partial bisimilariy coincide for nondeterministic automata.
\begin{lem}[$\PB = \PBn$]
\label{lem:if-det-S-det-P-then-PB-is-PBn}
Let $S$ and $P$ be deterministic automata.
Then $S \PB P$ iff $S \PBn P$.
\end{lem}

\begin{proof}
Immediate from the definitions.
\end{proof}

Rutten showed that on deterministic automata, partial bisimilarity is a control relation~\cite{rutten_coalgebra_2000}.

\begin{lem}[$\PB \subseteq \CR$]
\label{lem:if-det-S,det-P-then-PB-implies-CR}
Let $S$ and $P$ be deterministic automata. If $S \PB P$, then $S \CR P$.
\end{lem}

\begin{proof} 
Let
$S = \aut{S}$ and $P = \aut{P}$ be deterministic automata.
Let $R \subseteq \Q{S} \times \Q{P}$ be a partial bisimulation relation.
It follows immediately from the definitions that $R$ is also a control relation, so $S \CR P$.
\end{proof}

\begin{lem}[$\SC \subseteq \PB$]\label{lem:if-det-S,det-P-then-SC-implies-PB}
Let $S = \aut{S}$ and $P = \aut{P}$ be deterministic automata. If $S \SC P$, then $S \PB P$. 
\end{lem}

\begin{proof} 
Let $S$ and $P$ be such, and suppose $S \SC P$.
Let $f \colon \Q{S} \to \Q{P}$ be the embedding that witnesses the state controllability and define the relation $R = \{ (s, f(s)) \mid s \in \Q{S} \}$.
We show that $R$ is a partial bisimulation relation.
Pick arbitrary $q_S, q_P$ such that $q_S \R q_P$. By definition of $R$, $q_P = f(q_S)$.
We check the conditions of partial bisimilarity.
\begin{itemize}
  \item Let $a \in \Sigma$, $q'_S \in \Q{S}$ and $q_S \trans{a}{S} q'_S$. According to the definition of state controllability, $q_P = f(q_S) \trans{a}{P} f(q'_S)$, and by definition of $R$, $q'_S \R f(q'_S)$.
  \item Let $u \in \Sigmau$, $q'_P \in \Q{P}$ and $q_P \trans{u}{P} q'_P$.
  By definition of state controllability, there exists $q'_S$ such that $q_S \trans{u}{S} q'_S$. By the definition of state controllability, $q_P = f(q_S) \trans{u}{P} f(q'_S)$. Since $P$ is deterministic $f(q_S) = q'_P$, and since $q'_S \R f(q'_S) = q'_P$, the result follows. \qedhere
\end{itemize}
\end{proof}

\noindent
{\it Proof of Theorem~\ref{thm:det-S,det-P}.}
Theorem~\ref{thm:det-S-ndet-P} provides us with $\AC=\FM=\KT$, also for deterministic supervisors and plants. Hence, we do not consider $\FM$ and $\KT$ any further in the proofs below.
\begin{itemize}
    \item $\AC=\CR$ follows from Lemma~\ref{lem:if-det-S-det-P-then-AC-is-CR};
    \item $\PB = \PBn$ follows from Lemma~\ref{lem:if-det-S-det-P-then-PB-is-PBn}.
\end{itemize}
For the remaining proofs we only need to consider one notion from each equivalence class. We only consider $\SC$, $\PB$ and $\CR$.
\begin{itemize}
    \item $\PB \subseteq \CR$ follows from Lemma~\ref{lem:if-det-S,det-P-then-PB-implies-CR};
    \item $\SC \subseteq \PB$ follows from Lemma~\ref{lem:if-det-S,det-P-then-SC-implies-PB};
    \item $\PB \not\subseteq \SC$ follows from Example~\ref{exa:if-det-S,det-P-then-AC,FM,PB-does-not-imply-SC} since $\PB=\PBn$.
    \item Recall supervisor $S$ and plant $P$ from Example~\ref{exa:if-det-S,det-P-then-AC,FM-does-not-imply-PB}. Both are deterministic. As $S \LC P$ and $\LC = \CR$, $S \CR P$; from the example also $S \mathrel{\cancel{\PB}} P$, so combined with $\PB=\PBn$ this gives $\CR \not \subseteq \PB$.
    \qed
\end{itemize}

\subsection{Subautomata}

In this section, we provide the details required to establish the results in Theorem~\ref{thm:unlabeled}, capturing the relations between controllability notions for nondeterministic supervisors and plants, where the supervisor is a subautomaton of the plant.

\begin{lem}[$\FMsubseteq\subseteq \FM$]\label{lem:ndet-S,ndet-P,S-subseteq-P-then-FMsubseteq-implies-FM}
Let $S$ and $P$ be nondeterministic automata such that $S \subseteq P$.
If $S \FMsubseteq P$, then $S \FM P$.
\end{lem}
\begin{proof}
Immediate from the definition.
\end{proof}

\begin{lem}\label{lem:ndet-S,ndet-P,lang-S-subseteq-lang-P-then-SCn-is-FM}
Let $S$ and $P$ be nondeterministic automata such that $L(S) \subseteq L(P)$.
Then $S \SCn P$ iff $S \FM P$.
\end{lem}
\begin{proof}
Immediate from the definitions. As a side note, $\SCn = \KT$ is already shown in \cite{kushi_synthesis_2018}. Combined with $\FM = \KT$ this equality also proves the lemma.
\end{proof}

Since $S \subseteq P$ implies $L(S) \subseteq L(P)$, we also immediately have that Lemma~\ref{lem:ndet-S,ndet-P,lang-S-subseteq-lang-P-then-SCn-is-FM} also holds for the context discussed in this section.

\begin{lem}[$\FMsubseteq\subseteq\PBn$]\label{lem:ndet-S,ndet-P,S-subseteq-P-then-FMsubseteq-implies-PBn}
Let $S$ and $P$ be nondeterministic automata such that $S \subseteq P$.
If $S \FMsubseteq P$ then $S \PBn P$.
\end{lem}
\begin{proof} 
Let $S = \aut{S}$ and $P = \aut{P}$ be nondeterministic automata and suppose $S \FMsubseteq P$.
We define the relation $R = \{ (q_S, q_S) \in Q_S \times Q_P \mid q_S\in \mathit{Reach}(S)\}$. 
We show that $R$ is a nondeterministic partial bisimulation relation.
Pick arbitrary $q_S$ such that $q_S \R q_S$.
\begin{itemize}
    \item For the transfer condition from left to right, pick arbitrary $q'_S \in \Q{S}$, $a \in \Sigma$ such that $q_S \trans{a}{S} q'_S$. As $S \subseteq P$, $q'_S \in \Q{P}$, and $q_S \trans{a}{P} q'_S$, and by definition of $R$, $q'_S \R q'_S$.
    \item For the transfer condition from right to left, pick arbitrary $q'_P \in \Q{P}$, $u \in \Sigmau$ such that $q_S \trans{u}{P} q'_P$. As $q_S$ is reachable in $S$, and $S \subseteq P$, $q_S$ is reachable in $P$. Let $w \in \Sigma^*$ be a witness such that $\qinit{S} \trans{w}{S} q_S$ and $\qinit{S} \trans{w}{P} q_S$. So, we have  $\qinit{S} \trans{w}{P} q_S \trans{u}{P} q'_P$ and $\qinit{S} \trans{w}{S} q_S$. As $S \FMsubseteq P$ we then also have $q_S \trans{u}{S} q'_P$, so $q'_P \in \Q{S}$, hence $q'_P \R q'_P$. 
\end{itemize}
So, $R$ is a nondeterministic partial bisimulation relation, hence $S \PBn P$. \qedhere
\end{proof}

\noindent
{\it Proof of Theorem~\ref{thm:unlabeled}.}
Theorem~\ref{thm:nondet-S,nondet-P} provides us with $\FM = \KT$ and $\FM \subseteq \AC$;
$\SCn = \FM$ follows from Lemma~\ref{lem:ndet-S,ndet-P,lang-S-subseteq-lang-P-then-SCn-is-FM}. \footnote{Note that to obtain $\SCn = \FM$ it suffices that $L(S) \subseteq L(P)$.}
\begin{itemize}
 \item $\FMsubseteq \subsetneq \FM$ follows from Lemma~\ref{lem:ndet-S,ndet-P,S-subseteq-P-then-FMsubseteq-implies-FM} and Example~\ref{exa:ndet-S,ndet-P,S-subseteq-P-FMsubseteq};
 \item $\FMsubseteq \subsetneq \PBn$ follows from Lemma~\ref{lem:ndet-S,ndet-P,S-subseteq-P-then-FMsubseteq-implies-PBn} and Example~\ref{exa:ndet-S,ndet-P,S-subseteq-P-FMsubseteq};
 \item $\AC \not \subseteq \FM$ follows from Example~\ref{exa:ndet-S,ndet-P,S-subseteq-P-FM-vs-AC};
 \item $\PBn \not \subseteq \AC$ follows from Example~\ref{exa:if-det-S-nondet-P-then-PB-does-not-imply-FM-AC} provided that $s_1$, $s_2$ and $s_3$ are renamed into $p_1$, $p_2$ and $p_3$, respectively, in the supervisor (for it to be a subautomaton of $P$); 
 \item $\AC \not \subseteq \PBn$ follows from Example~\ref{exa:det-S,ndet-P,S-subseteq-P,AC,FM-do-not-imply-PBn,FMsubseteq};
 \item $\PBn \not \subseteq \FM$ follows from $\FM \subseteq \AC$ and $\PBn \not \subseteq \AC$. $\FM \not \subseteq \PBn$ follows from Example~\ref{exa:det-S,ndet-P,S-subseteq-P,AC,FM-do-not-imply-PBn,FMsubseteq}.
 \qed
\end{itemize}

\begin{exmp}\label{exa:ndet-S,ndet-P,S-subseteq-P-FMsubseteq}
Consider the supervisor $S$ and plant $P$ with $S \subseteq P$, and $\Sigmau = \{ u \}$ as depicted below.
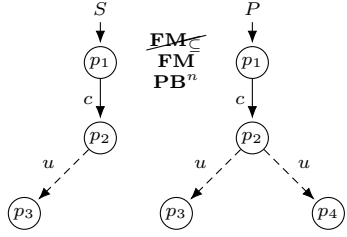
\begin{center} 
\begin{tikzpicture}[>=Latex,shorten >=1pt,auto,->,node distance=1cm,align=center,font=\it, uncontrollable/.style={densely dashed}]
\scriptsize
\node[initial above, initial text = {$S$}, state] (s1) {$p_1$};
\node[state, below of=s1] (s2) {$p_2$};
\node[state, below of=s2, left of = s2] (s3) {$p_3$};

\path[->]
    (s1) edge node[left] {$c$} (s2)
    (s2) edge[uncontrollable] node[above left] {$u$} (s3)
    ;

\node[right of = s1] (c) {$\begin{array}{c}\cancel{\FMsubseteq}\\\FM\\\PBn\end{array}$};

\node[initial above, initial text = {$P$}, state, right of=c] (p1) {$p_1$};
\node[state, below of=p1] (p2) {$p_2$};
\node[state, below of=p2, left of=p2] (p3) {$p_3$};
\node[state, below of=p2, right of=p2] (p4) {$p_4$};

\path[->]
    (p1) edge node[left] {$c$} (p2)
    (p2) edge[uncontrollable] node[above left] {$u$} (p3)
    (p2) edge[uncontrollable] node[above right] {$u$} (p4)
    ;

\end{tikzpicture}
\end{center}

$S \FM P$ follows from the fact that for the only sequence of events in $P$ that ends with an uncontrollable event, i.e., $p_1 \trans{cu}{P}$, and the only transition with $c$ in $S$, i.e., $p_1 \trans{c}{S} p_2$, we also have $p_2 \trans{u}{S}$. We cannot have $S \FMsubseteq P$, since the sequence $p_1 \trans{c}{P} p_2 \trans{u}{P} p_4$ and the sequence $p_1 \trans{c}{S} p_2$ are present but not $p_2 \trans{u}{S} p_4$, as required by $\FMsubseteq$. Note that we also have $S \PBn P$ as witnessed by the nondeterministic partial bisimulation $\{ (p_1,p_1), (p_2,p_2),(p_3,p_3), (p_3,p_4) \}$.
\end{exmp}

\begin{exmp}\label{exa:ndet-S,ndet-P,S-subseteq-P-FM-vs-AC}
Consider the plant $P$ with $\Sigmau = \{ u \}$ as depicted below, and a supervisor $S = P$.
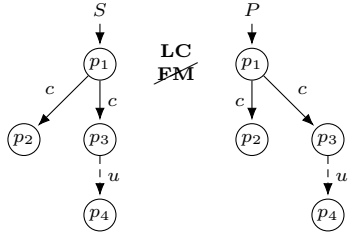
\begin{center} 
\begin{tikzpicture}[>=Latex,shorten >=1pt,auto,->,node distance=1cm,align=center,font=\it, uncontrollable/.style={densely dashed}]
\scriptsize
\node[initial above, initial text = {$S$}, state] (s1) {$p_1$};
\node[state, below of=s1, left of=s1] (s2) {$p_2$};
\node[state, below of=s1] (s3) {$p_3$};
\node[state, below of=s3] (s4) {$p_4$};

\path[->]
    (s1) edge node[above left] {$c$} (s2)
    (s1) edge node[right] {$c$} (s3)
    (s3) edge[uncontrollable] node[right] {$u$} (s4)
    ;

\node[right of = s1] (c) {$\begin{array}{c}\LC\\\cancel{\FM}\end{array}$};

\node[initial above, initial text = {$P$}, state, right of=c] (p1) {$p_1$};
\node[state, below of=p1] (p2) {$p_2$};
\node[state, below of=p1, right of=p1] (p3) {$p_3$};
\node[state, below of=p3] (p4) {$p_4$};

\path[->]
    (p1) edge node[left] {$c$} (p2)
    (p1) edge node[above right] {$c$} (p3)
    (p3) edge[uncontrollable] node[right] {$u$} (p4)
    ;
\end{tikzpicture}
\end{center}
Note that trivially $S \subseteq P$, and that $S \LC P$. However, $S \mathrel{\cancel{\FM}} P$, because $p_1 \trans{cu}{P}$ and $p_1 \trans{c}{S} p_2$, but not $p_2 \trans{u}{S}$.
\end{exmp}

\begin{exmp}\label{exa:det-S,ndet-P,S-subseteq-P,AC,FM-do-not-imply-PBn,FMsubseteq}
Consider the supervisor $S$ and plant $P$ with $S \subseteq P$, and $\Sigmau = \{ u \}$ as depicted below.
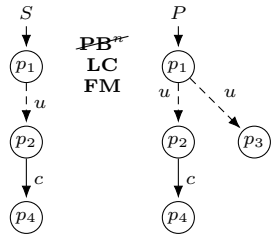
\begin{center} 
\begin{tikzpicture}[>=Latex,shorten >=1pt,auto,->,node distance=1cm,align=center,font=\it, uncontrollable/.style={densely dashed}]
\scriptsize
\node[initial above, initial text = {$S$},state] (s1) {$p_1$};
\node[state, below of=s1] (s2) {$p_2$};
\node[state, below of=s2] (s4) {$p_4$};

\path[->]
    (s1) edge[uncontrollable] node[right] {$u$} (s2)
    (s2) edge node[right] {$c$} (s4)
    ;

\node[right of = s1] (c) {$\begin{array}{c}\cancel{\PBn}\\
\LC\\
\FM
\end{array}$};

\node[initial above, initial text = {$P$},state, right of=c] (p1) {$p_1$};
\node[state, below of=p1] (p2) {$p_2$};
\node[state, below of=p1, right of=p1] (p3) {$p_3$};
\node[state, below of=p2] (p4)  {$p_4$};

\path[->]
    (p1) edge[uncontrollable] node[above left] {$u$} (p2)
    (p1) edge[uncontrollable] node[above right] {$u$} (p3)
    (p2) edge  node[right]  {$c$} (p4)
    ;

\end{tikzpicture}
\end{center}
$S \FM P$ follows from the fact that $u$ is the only sequence of events in $P$ that ends in an uncontrollable event, i.e., $p_1 \trans{u}{P}$, in $S$, we also have $p_1 \trans{u}{S} p_2$. $S \LC P$ follows from this and $\FM \subseteq \LC$.
However, every nondeterministic partial bisimulation relation must relate $p_1$ in $S$ and $P$, which means that the transition $p_1 \trans{u}{P}  p_3$ must be mimicked by some transition in $S$. Such transition does not exist.
\end{exmp}

\end{document}